%% file: main.tex
\documentclass{article}

 \PassOptionsToPackage{numbers, compress}{natbib}
\usepackage[preprint]{neurips_2023}

\usepackage[utf8]{inputenc} % allow utf-8 input
\usepackage[T1]{fontenc}    % use 8-bit T1 fonts
\usepackage{url}            % simple URL typesetting
\usepackage{booktabs}       % professional-quality tables
\usepackage{amsfonts}       % blackboard math symbols
\usepackage{nicefrac}       % compact symbols for 1/2, etc.
\usepackage{microtype}      % microtypography
\usepackage{xcolor}         % colors

\usepackage{algorithmic}
\usepackage{graphicx}
\usepackage{textcomp}
\usepackage{enumitem}
\usepackage{multirow}
\usepackage{makecell}

\input{math_commands}

\newcommand{\eg}{\textit{e}.\textit{g}., }
\newcommand{\ie}{\textit{i}.\textit{e}., }

\newcommand{\plg}{P\textsuperscript{13}LG }
\newcounter{bxincomm}
\definecolor{aqua}{rgb}{0.00,0.67,0.80}

\newcounter{ygincomm}

\newcounter{todocomm}

\title{Multi-level Protein Representation Learning for \\ Blind Mutational Effect Prediction
}

\author{%
  Yang Tan $^*$\\
  East China University of Science and Technology \\
  Shanghai, China, 200237 \\
  \texttt{tyang@mail.ecust.edu.cn} \\
  \And
  Bingxin Zhou \thanks{equal contribution.} \\
  Shanghai Jiao Tong University \\
  Shanghai, China, 200240 \\
  \texttt{bingxin.zhou@sjtu.edu.cn} \\
  \AND
  Yuanhong Jiang\\
  Shanghai Jiao Tong University \\
  Shanghai, China, 200240 \\
  \texttt{william\_jiang@sjtu.edu.cn} \\
  \And
  Yu Guang Wang \\
  Shanghai Jiao Tong University \\
  Shanghai, China, 200240 \\
  \texttt{yuguang.wang@sjtu.edu.cn} \\
  \And
  Liang Hong \\
  Shanghai Jiao Tong University \\
  Shanghai, China, 200240 \\
  \texttt{hongl3liang@sjtu.edu.cn} \\
}

\begin{document}

\maketitle

\begin{abstract}
  Directed evolution plays an indispensable role in protein engineering that revises existing protein sequences to attain new or enhanced functions. Accurately predicting the effects of protein variants necessitates an in-depth understanding of protein structure and function. Although large self-supervised language models have demonstrated remarkable performance in zero-shot inference using only protein sequences, these models inherently do not interpret the spatial characteristics of protein structures, which are crucial for comprehending protein folding stability and internal molecular interactions. This paper introduces a novel pre-training framework that cascades sequential and geometric analyzers for protein primary and tertiary structures. It guides mutational directions toward desired traits by simulating natural selection on wild-type proteins and evaluates the effects of variants based on their fitness to perform the function. We assess the proposed approach using a public database and two new databases for a variety of variant effect prediction tasks, which encompass a diverse set of proteins and assays from different taxa. The prediction results achieve state-of-the-art performance over other zero-shot learning methods for both single-site mutations and deep mutations.
\end{abstract}

\section{Introduction}

The analysis of protein sequence–function relationships provides valuable insights for enzyme engineering to develop new or enhanced functions. Predicting the effects of point mutations in proteins allow researchers to dissect how changes in the amino acid (AA) sequence can impact the protein's structure, stability, function, and interactions with other molecules \cite{shin2021protein}. While direct projection to protein functionality may encompass numerous uncharacterized molecular interactions, the advent of high-throughput experimental techniques has enabled the measurement of sequence-function mappings, thereby expanding the range of observable biochemical functions \cite{fowler2014deep, melamed2013deep, podgornaia2015pervasive, romero2015dissecting}. Currently, hundreds of well-studied proteins have documented tens of thousands of mutational effects on their functions, such as ParD-ParE complexes for binding preferences \cite{aakre2015evolving}, and ubiquitin for thermodynamic stability \cite{roscoe2013analyses}, and green fluorescent proteins for fluorescence \cite{sarkisyan2016local}, to name but a few. 
Systematic exploration of sequence variants offers copious evidence for characterizing the evolutionary space of mutants. However, this approach heavily depends on the domain-specific knowledge of individual proteins and their functionalities, thereby generalizing these specifically-designed mapping rules to a vast array of protein families poses a significant challenge.

Deep learning methods have been applied to expedite protein research, particularly in bridging protein sequences to functions. Analogous to how language models analyze the semantics and syntax of human language, these methods interpret protein sequences as raw text and utilize self-attention mechanisms \cite{brandes2022proteinbert,nijkamp2022progen2,rives2021esm1b} and/or autoregressive inference methods \cite{madani2023progen,notin2022tranception} to reveal hidden long-range sequential dependencies. In addition, multiple sequence alignments (MSAs) have been wildly applied in predicting protein sequence \cite{ahmed2021prottrans,hsu2022esmif,meier2021esm1v,rao2021msa} or structure \cite{baek2021accurate,jumper2021alphafold} to augment the contextual information gleaned from sets of relevant sequences. 
While language models reveal sophisticated projections from protein sequence to functionality, inferring hundreds to thousands of uncharacterized molecular interactions demands considerable input samples and computationally intensive propagation modules. Alternatively, the structure of a protein, governed by its AA sequence, not only guides molecular interactions but also determines its functionality. Studies have derived the latent representation of AAs' local environment based on protein topology or geometry \cite{jones2021improved,torng2019high,zhou2023accurate}. They assume that an accurate description of a protein's microenvironment is essential for determining its properties or functionality. Given that core mutations often induce functional defects through subtle disruptions to structure or dynamics \cite{roscoe2013analyses}, incorporating protein topology or geometry into the learning process can offer valuable insights into stabilizing protein functioning.

Both sequence and structure-oriented deep learning approaches have contributed to significant scientific discoveries \cite{lu2022machine,madani2023progen,shroff2020discovery}. However, they exhibit limitations when implemented individually. Structure encoders fail to capture long-range sequential connections for an AA beyond its contact region, and they overlook any correlations that do not conform to the `structure-function determination' heuristic. On the other hand, sequential encoders struggle to capture the spatial molecular interactions of long-range elements and require an excessive number of protein instances to implicitly unravel the deep connection between protein topology and functionality. Although sequence-based learning might or might not find a better solution than human beings, language models demand significantly more resources for data processing and model training. In natural language processing, large models claim to consume more than $10^{12}$ documents to achieve quantitative changes in inference \cite{rasley2020deepspeed,shoeybi2019megatron}.

\begin{figure}[!t]
    \centering
    \includegraphics[width=\textwidth]{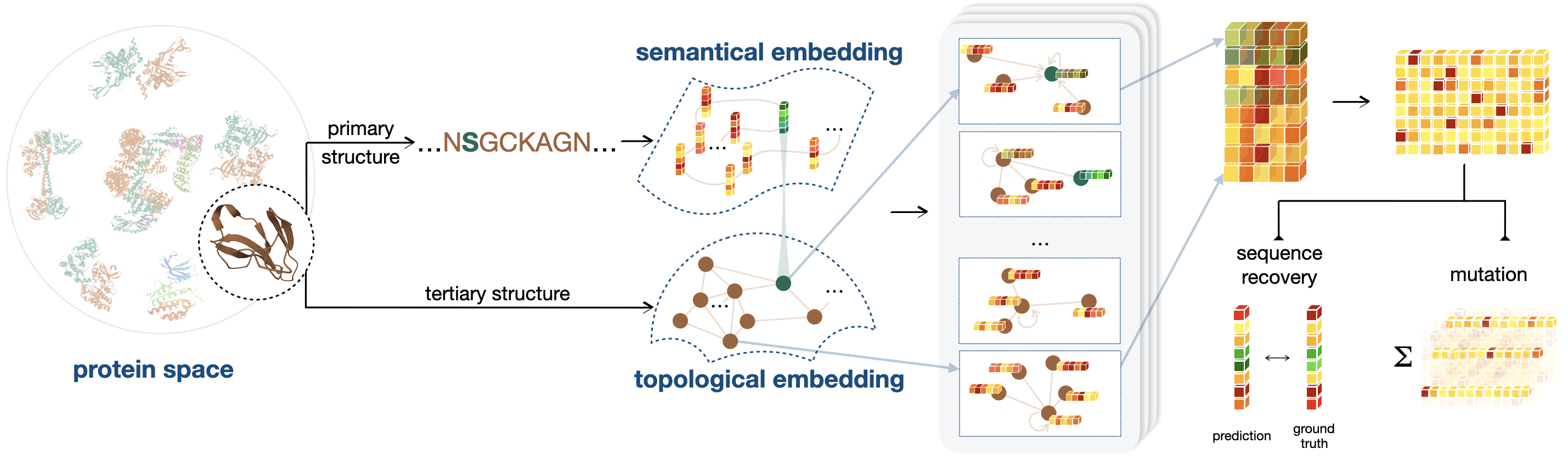}
    \caption{An illustration of \plg that extracts the semantics and topology of a protein by learning its primary and tertiary structures. The hidden representation can be decoded for variants effect prediction that recognizes the impact of mutating a few sites of a protein on its functionality.}
    \label{fig:main_architecture}
\end{figure}

We believe incorporating the intermediate state of protein structures can facilitate the discovery of an efficient and effective trajectory for mapping protein sequences to functionalities. To this end, we introduce \plg, a framework designed to assimilate the semantics and topology of \textbf{\underline{P}}roteins from their primary (\textbf{\underline{1}}) and tertiary (\textbf{\underline{3}}) structure with \textbf{\underline{L}}anguage and \textbf{\underline{G}}raph models. The developed model extends the generalization and robustness of self-supervised protein language models while maintaining low computational costs, thereby facilitating self-supervised training and task-specific customization. A funnel-shaped learning pipeline, as depicted in Figure~\ref{fig:main_architecture}, is designed due to the limited availability of crystallized protein structures compared to observed protein sequences. Initially, the linguistic embedding establishes the semantic and grammatical rules in AA chains by inspecting over $60$ million protein sequences \cite{lin2023esm2}. Then, the topological embedding encodes the microenvironment of AAs, supplementing sequential relationships with spatial connections. Since a geometry can be placed or observed by different angles and positions in space, we represent proteins' topology by graphs and enhance the model's robustness and efficiency with a rotation and translation equivariant graph representation learning scheme. Consequently, the trained model is capable of interpreting the characterization of proteins in dimensions closely related to their functionality. 

The developed model for directed evolution fulfills practical requirements in enzyme engineering from three perspectives. (i) The model gains interpretability by simulating natural selection. During training, random perturbations are assigned to AA types to encourage the model to recover advantageous protein sequences found in nature. (ii) The trained model provides robust and meaningful approximations to the joint distribution of the complete AA chain, enhancing the \textit{epistatic effect} \cite{khersonsky2018automated,sarkisyan2016local} in deep mutations by considering the nonlinear combinatorial effects of AA sites. (iii) The model deploys self-supervised learning during training to eliminate the need for further supervision on downstream tasks. This zero-shot scenario is desirable due to the scarcity of experimental results as well as the `cold-start' situation common in many wet lab experiments.

The pre-trained \plg demonstrates its feasibility across a broad range of variant effect prediction benchmarks. These benchmarks include a general deep mutational effect prediction benchmark, \textbf{ProteinGym} \cite{notin2022tranception}, which comprises over $80$ proteins of varying assays and taxa. In addition, we have prepared two niche single-site mutation benchmarks.  They measure thermodynamic stability using $\Delta$Tm and $\Delta\Delta$G values and $2,967$ mutants across $90$ protein-condition combinations. These two databases supplement the existing publicly available benchmarks with assay-specific deep mutational scanning (DMS) records, which facilitate the establishment of well-defined evaluation criteria for future methods that are specifically designed and assessed based on protein stability.

\section{Zero-shot Multi-level Protein Representation Learning}
\label{sec:dataPrep}
Labeled data are usually scarce in biomolecule research, which demands designing a general model for predicting variant effects on unknown proteins and protein functions.  Given a three-dimensional protein backbone structure, this study utilizes a self-supervised learning model that recovers AA types from noisy local environment observations. It simulates nature's iterative selection of well-functioning mutated proteins from harmful random mutations.

\subsection{Multi-level Protein Representation}
\paragraph{Protein Primary Structure (Noised)}
 For a given observation with an AA type $\tilde{\vv}$, it is assumed that this observation is randomly perturbed. The model then learns a revised state $\vv$ that is less likely to be eliminated by natural selection due to unfavorable properties such as instability or inability to fold. Formally, we define the perturbed observation by a Bernoulli distribution as follows:
\begin{equation}
\label{eq:noise_aa}
    \boldsymbol{\pi}(\tilde{\vv} \mid \vv)=p\boldsymbol{\Theta}(\pi_1,\pi_2,\dots,\pi_{20})+(1-p)\delta(\tilde{\vv}-\vv),
\end{equation}
where an AA in a protein chain has a chance of $p$ to mutate to one of $20$ AAs following the \emph{replacement distribution} $\boldsymbol{\Theta}(\cdot)$ and $(1-p)$ of remaining unchanged. We consider $p$ as a tunable parameter and define $\boldsymbol{\Theta}(\cdot)$ based on the frequency of AA types observed in wild-type proteins in the training dataset.

\paragraph{Protein Tertiary Structure}
The geometry of a protein is described by $\gG=(\gV,\gE, \mW_V, \mW_E, \mX_V)$, a residue graph %\cite{xiong2023graph} 
by the $k$-nearest neighbor ($k$NN). Each node $v_i\in\gV$ represents an AA in the protein connected to up to $k$ other nodes in the graph that are the closest in Euclidean distance within a contact region of $30$\AA. Node attributes $\mW_V$ are hidden semantic embeddings of AA types, and edge attributes $\mW_E\in\R^{93}$ feature relationships of connected nodes based on inter-atomic distances, local N-C positions, and sequential position encoding. Additionally, $\mX_V$ records 3D coordinates of AAs in the Euclidean space, which plays a crucial role in the subsequent topological embedding stage to preserve roto-translation equivariance.

\subsection{Semantic Encoding of Protein Sequence}
Although it is generally believed by the community that a protein's sequence determines its biological function via the folded structure, following strictly to this singular pathway risks overlooking other unobserved yet potentially influential inter-atomic communications impacting protein fitness. In line with this reasoning, our proposed model, \plg, begins by extracting pairwise relationships for residues through an analysis of proteins' primary structure from $\tilde{\mV}$ and embed them to hidden representations $\mW_V$ for residues. At each update, the information and representations of the noisy AA sequence are encoded from the noisy input via an \textit{Evolutionary Scale Modeling} \textsc{ESM-2} \cite{lin2023esm2}\footnote{Official implementation released at \url{https://github.com/facebookresearch/esm}.}. This approach employs a BERT-style masked language modeling (MLM) objective that predicts the identity of randomly selected AAs in a protein sequence by observing their context within the remainder of the sequence. Note that during training, the sequence embedding operates in every epoch as AA types are subject to independent random perturbations. For alternative encoding strategies, please refer to the discussion in Appendix~\ref{sec:app:noise}.

\subsection{Topological Encoding of Protein Structure}
Proteins are structured in 3D space, which requires the geometric encoder to possess roto-translation equivariance to node positions as well as permutation invariant to node attributes. This design is vital to avoid the implementation of costly data augmentation strategies. We practice \emph{Equivariant Graph Neural Networks} (EGNN) \cite{satorras2021n} to acquire the hidden representation to node properties $\mW_V^{l+1}=\left\{\vw_{v_1}^{l+1}, \dots, \vw_{v_n}^{l+1}\right\}$ and node coordinates $\mX_{\text{pos}}^{l+1}=\left\{\vx_{v_1}^{l+1}, \dots, \vx_{v_n}^{l+1}\right\}$ at the $l+1$th layer by 
\begin{equation}
\label{eq:egnn}
    \begin{aligned}
    \vm_{ij} &=\phi_{e}\left(\vw_{v_i}^{l}, \vw_{v_j}^{l},\left\|\vx_{v_i}^{l}-\vx_{v_j}^{l}\right\|^{2}, \vw_{e_{ij}}\right), \\
    \vx_{v_i}^{l+1} &=\mathbf{x}_{v_i}^{l}+\frac1{n}\sum_{j \neq i}\left(\mathbf{x}_{v_i}^{l}-\mathbf{x}_{v_j}^{l}\right) \phi_{x}\left(\mathbf{m}_{{ij}}\right), \quad \\
    \vw_{v_i}^{l+1} &=\phi_{v}\big(\vw_{i}^{l}, \sum_{j \neq i} \vm_{i j}\big).
\end{aligned}
\end{equation}
In these equations, $\vw_{e_{ij}}$ represents the input edge attribute on $\gV_{ij}$, which is not updated by the network. The propagation rules $\phi_{e}, \phi_{x}$ and $\phi_{v}$ are defined by differentiable functions, \eg multi-layer perceptrons (MLPs). The final hidden representation on nodes $\mW_V^{L}$ embeds the microenvironment and local topology of AAs, and it will be carried on by readout layers for label predictions.

\section{Blind Variant Effect Prediction}
Our method is specifically designed for protein engineering that is trained with a self-supervised learning scheme.
The model's capability extends to blind variant effect prediction on an unknown protein, and it can generate the joint distribution for all AA sites as one of the 20 possible types, conditioned on their spatial and sequential neighbors. This process accounts for the epistatic effect and concurrently returns all AA sites in a sequence. Below we detail the workflow for training the zero-shot model and scoring the mutational effect of a specific mutant.

\subsection{Model Pipeline}
\paragraph{Training}
The fundamental model architecture cascades a frozen sequence encoding module and a trainable tertiary structure encoder. Initially, a protein language model encodes pairwise hidden relationships of AAs by analyzing the input protein sequence and produces a vector representation $\vw_{v_i}\in\mW_V$ for an arbitrary AA, where $\mW_V=\text{LM}_{\text{frozen}}(\tilde{\mV})$ with $\tilde{\mV}$ be the perturbed initial AA-type encoding. The language model $\text{LM}_{\text{frozen}}(\cdot)$, \textsc{ESM-2} \cite{lin2023esm2} for instance, has been pre-trained on a massive protein sequence database (\eg \textbf{UniRef50} \cite{suzek2015uniref}) to understand the semantic and grammatical rules of wild-type proteins. It conceals high-dimensional AA-level long-short-range interactions that may or may not have been investigated and explained by scientists. Next, we represent proteins by $k$NN graphs with model-encoded node attributes, handcrafted edge attributes, and 3D positions of the corresponding AAs. This representation is embedded using a stack of $L$ \textsc{EGNN} \cite{satorras2021n} layers to yield $\mW_V^L=\text{EGNN}(\gG)$. This process extracts the geometric and topological embedding for protein graphs with AAs represented by $\vw_{v_i}$. During the pre-training phase for protein sequence recovery, the output layer $\phi(\cdot)$ provides the probability of AA types on each residue, \ie $\mY=\phi(\mW_V^{L})\in\R^{n\times 20}$ for a protein comprising $n$ AAs. The model's learnable parameters are refined by minimizing the cross-entropy of the recovered AAs with respect to the ground-truth AAs in wild-type proteins.

\paragraph{Inference}
For a given mutant, its fitness score is derived from the joint distribution of the altered AA types on associated nodes that provides a preliminary assessment based on natural observations. We consider the AA type in the wild-type protein as a reference state and compare it with the predicted probability of AAs at the mutated site. Formally, for a mutant with mutated sites $\gT$ ($|\gT|\geq 1$), we define its fitness score by the corresponding \textit{log-odds-ratio}, \ie
$\sum_{t\in\gT}\log p(\vy_t)-\log p(\vv_t)$, where $\vy_t$ and $\vv_t$ denote the mutated and the wild-type AA at site $t$, respectively.

\subsection{Evaluation Metrics}
It is critical to evaluate the developed model's effectiveness using quantitative, computable measurements before proceeding to wet lab validations. Within the scope of mutational effect prediction, each raw protein in the database maintains dozens to tens of thousands of mutants with varying depths of mutation sites. Considering that protein functions are sensitive to the external environment and experimental methods, the absolute values measured by individual labs are typically not directly comparable. Consequently, we evaluate the performance of pre-trained models on a diverse set of proteins and protein functions using two quantitative measurements for ordinal and categorical data.

\paragraph{Spearman's $\rho$ Correlation}
Spearman's correlation is commonly applied in mutational effect prediction tasks to measure the strength and direction of the monotonic relationship between two ranked sequences \ie experimentally evaluated mutants and model-inferred mutants. This non-parametric rank measure is robust to outliers and asymmetry in mutational scores, does not assume any specific distribution of mutational scores, and captures non-linear correlations between the two sequences. The scale of $\rho$ ranges from $-1$ to $1$ which indicates whether the predicted sequence is negatively or positively related to the ground truth. Ideally, a result close to $1$ is preferred.

\paragraph{True Positive Rate}
The true positive rate (TPR), also known as recall, is a key performance measure for the proportion of actual positives that are correctly identified. In the context of directed evolution tasks, this refers to the proportion of top beneficial mutations that are accurately predicted as most beneficial. A high TPR for the predicted results indicates that the trained model is likely to provide reliable mutational recommendations for wet labs. The following section will test TPR for baseline models on each of the proteins at $5\%$, $25\%$, and $50\%$. For instance, TPR at $5\%$ defines the top $5\%$ mutants (in terms of the highest ground-truth score) as `positive samples' and measures the proportion of these samples that are also ranked in the top $5\%$ by the model.

\section{Numerical Experiments}
\label{sec:experiment}
We validate the efficacy of \plg on zero-shot mutational effect prediction tasks on $186$ diverse proteins. The performance is compared with other SOTA models of varying scales (\ie number of parameters). The implementations (\url{https://anonymous.4open.science/r/plg-1B02}) are programmed with \texttt{PyTorch-Geometric} (ver 2.2.0) and \texttt{PyTorch} (ver 1.12.1) and executed on an NVIDIA$^{\circledR}$ Tesla A100 GPU with $6,912$ CUDA cores and $80$GB HBM2 installed on an HPC cluster.

\subsection{Experimental Protocol}

\paragraph{Training Setup}
We train \plg on a non-redundant subset of \textbf{CATH v4.3.0} \cite{ORENGO19971093} domains, which contains $30,948$ experimental protein structures with less than $40$\% sequence identity. We further remove $\sim6\%$ of proteins that exceed $2,000$ AAs in length. Each protein domain is transformed into a $k$NN graph as following Section~\ref{sec:dataPrep}, with node features extracted by a frozen \textsc{ESM2}-t33 \cite{lin2023esm2} prefix model. Protein topology is inferred by a $6$-layer EGNN \cite{satorras2021n} with the hidden dimension tuned from $\{512,768,1280\}$. \textsc{Adam} \citep{kingma2014adam} is used for backpropagation with the learning rate set to $0.0001$. To avoid training instability or CUDA out-of-memory errors, we limit the maximum input to $8,192$ AA tokens per batch, constituting approximately $32$ residue graphs.

\paragraph{Baseline Methods}
We undertake an extensive comparison with baseline methods of self-supervised SOTA models on the fitness of mutation effects prediction. These methods utilize protein sequences and/or structures for learning. Sequence models employ position embedding strategies such as autoregression (\textsc{Tranception} \cite{notin2022tranception}, \textsc{RITA} \cite{hesslow2022rita}, and \textsc{ProGen2} \cite{nijkamp2022progen2}), masked language modeling (\textsc{ESM-1b} \cite{rives2021esm1b}, \textsc{ESM-1v} \cite{meier2021esm1v}, and \textsc{ESM2} \cite{lin2023esm2}), and a combination of the both (\textsc{ProtTrans} \cite{ahmed2021prottrans}). As our model acquires structural encoding, we also compare with \textsc{ESM-IF1} \cite{hsu2022esmif} which incorporates mask language modeling objectives with \textsc{GVP} \cite{jing2020gvp}. \textbf{ProteinGym} exhibits diverse protein types and assays, we thus include additional baselines that utilize 
MSA for model training (\textsc{DeepSequence} \cite{riesselman2018deepsequence}, \textsc{WaveNet} \cite{shin2021wavenet}, 
\textsc{MSA-Transformer} \cite{rao2021msa}, \textsc{SiteIndep}, and \textsc{EVmutation} \cite{hopf2017evmutation}).

\paragraph{Benchmark Datasets}
We conduct a comprehensive comparison of diverse mutation effect predictors in different regimes. Following \cite{notin2022tranception,riesselman2018deepsequence}, we prioritize experimentally-measured properties that possess a monotonic relationship with protein fitness, such as protein stability and relative activity. For protein stability, we generate $90$ experimentally-measured sets of protein-condition combination assays from \textbf{ProThermDB} \footnote{Retrieved from \url{https://web.iitm.ac.in/bioinfo2/prothermdb/index.html}.}, containing $2,967$ single-site mutants in environments with different pH levels, where $60$ of them are measured by $\Delta$Tm (the change of melting temperature) and the rest $30$ assays are by $\Delta\Delta$G (the change in the change in Gibbs free energy). The two datasets are named according to their scoring metrics: \textbf{DTm} and \textbf{DDG}, respectively. See Appendix~\ref{sec:app:Tm_DDG} for additional descriptions. We also examine the fitness prediction of the proteins in \textbf{ProteinGym}, which constitutes $86$~\footnote{We exclude \texttt{A0A140D2T1\_ZIKV\_Sourisseau\_growth\_2019}, the longest protein of over $3,000$ AAs in \textbf{ProteinGym} because it fails to be folded by \textsc{AlphaFold2}.} DMS assays of different taxa (\eg prokaryotes, humans, other eukaryotes, viruses).

\begin{table}[!t]
\caption{Variant Effect Prediction on \textbf{DTm} and \textbf{DDG}.}
\label{tab:TM_DMS}
\begin{center}
\resizebox{0.7\textwidth}{!}{
    \begin{tabular}{cccccccc}
    \toprule
    \multirow{2}{*}{\textbf{Model}} & \multirow{2}{*}{\textbf{version}} & \multicolumn{3}{c}{\textbf{TPR $\uparrow$} (DTm)} & \multicolumn{3}{c}{\textbf{TPR $\uparrow$} (DDG) } \\\cmidrule(lr){3-5}\cmidrule(lr){6-8}
    &  & 5\% & 25\% & 50\% & 5\% & 25\% & 50\% \\
    \midrule
    \multirow{6}{*}{\textsc{ProGen2}} 
    & oas & 0.033 & 0.286 & 0.537 & 0.000 & 0.339 & 0.515 \\
    & medium & 0.117 & 0.367 & 0.582 & 0.072 & 0.443 & 0.615 \\
    & base & 0.212 & 0.362 & 0.585 & \textbf{0.231} & 0.408 & 0.621 \\
    & large & 0.132 & 0.323 & 0.557 & 0.117 & 0.320 & 0.597 \\
    & BFD90 & 0.178 & 0.333 & 0.589 & 0.206 & 0.451 & \textbf{0.644} \\
    & xlarge & 0.118 & 0.353 & 0.578 & 0.144 & 0.383 & 0.603 \\
    \midrule
    \multirow{2}{*}{\textsc{Tranception}} 
    & medium & 0.188 & 0.359 & 0.564 & 0.083 & 0.367 & 0.527 \\
    & large & 0.149 & 0.371 & 0.586 & 0.072 & 0.395 & 0.540 \\
    \midrule
    \multirow{4}{*}{\textsc{PortTrans}} 
    & bert & 0.131 & 0.364 & 0.586 & 0.122 & 0.424 & 0.635 \\
    & bert\_bfd & 0.168 & 0.336 & 0.579 & 0.136 & 0.423 & 0.589 \\
    & t5\_xl\_uniref50 & 0.184 & 0.412 & 0.593 & 0.147 & 0.425 & 0.640 \\
    & t5\_xl\_bfd & 0.136 & 0.350 & 0.587 & 0.106 & 0.419 & 0.610 \\
    \midrule
    \textsc{ESM-1v} 
    & -  & 0.216 & 0.386 & 0.602 & \textbf{0.231} & 0.451 & 0.622 \\
    \midrule
    \textsc{ESM-1b} 
    & - & 0.151 & 0.402 & 0.606 & 0.211 & 0.424 & 0.642\\
    \midrule
    \textsc{ESM-if1} 
    & -  & 0.188 & \textbf{0.418} & \textcolor{red}{\textbf{0.656}} & \textcolor{violet}{\textbf{0.258}} & \textcolor{red}{\textbf{0.469}} & 0.641 \\
    \midrule
    \multirow{4}{*}{\textsc{ESM-2}} 
    & t30 & 0.139 & 0.397 & 0.598 & 0.172 & \textbf{0.453} & \textcolor{violet}{\textbf{0.646}} \\
    & t33 & \textcolor{violet}{\textbf{0.239}} & 0.407 & 0.601 & 0.181 & 0.438 & 0.637 \\
    & t36 & 0.152 & 0.408 & \textbf{0.634}  & 0.169 & 0.405 & 0.641\\
    & t48 & \textbf{0.232} & \textcolor{red}{\textbf{0.430}} & 0.607 & 0.189 & 0.400 & 0.606 \\
    \midrule
    \plg 
    & k20\_h1280 & \textcolor{red}{\textbf{0.304}} & \textcolor{violet}{\textbf{0.419}} & \textcolor{violet}{\textbf{0.642}}  & \textcolor{red}{\textbf{0.267}} & \textcolor{violet}{\textbf{0.454}} & \textcolor{red}{\textbf{0.676}}\\
    \bottomrule\\[-2.5mm]
    \multicolumn{8}{l}{$\dagger$ The top three are highlighted by \textbf{\textcolor{red}{First}}, \textbf{\textcolor{violet}{Second}}, \textbf{Third}.}
    \end{tabular}
}
\end{center}
\end{table}

\begin{figure}[!t]
    \centering
    \includegraphics[width=0.8\textwidth]{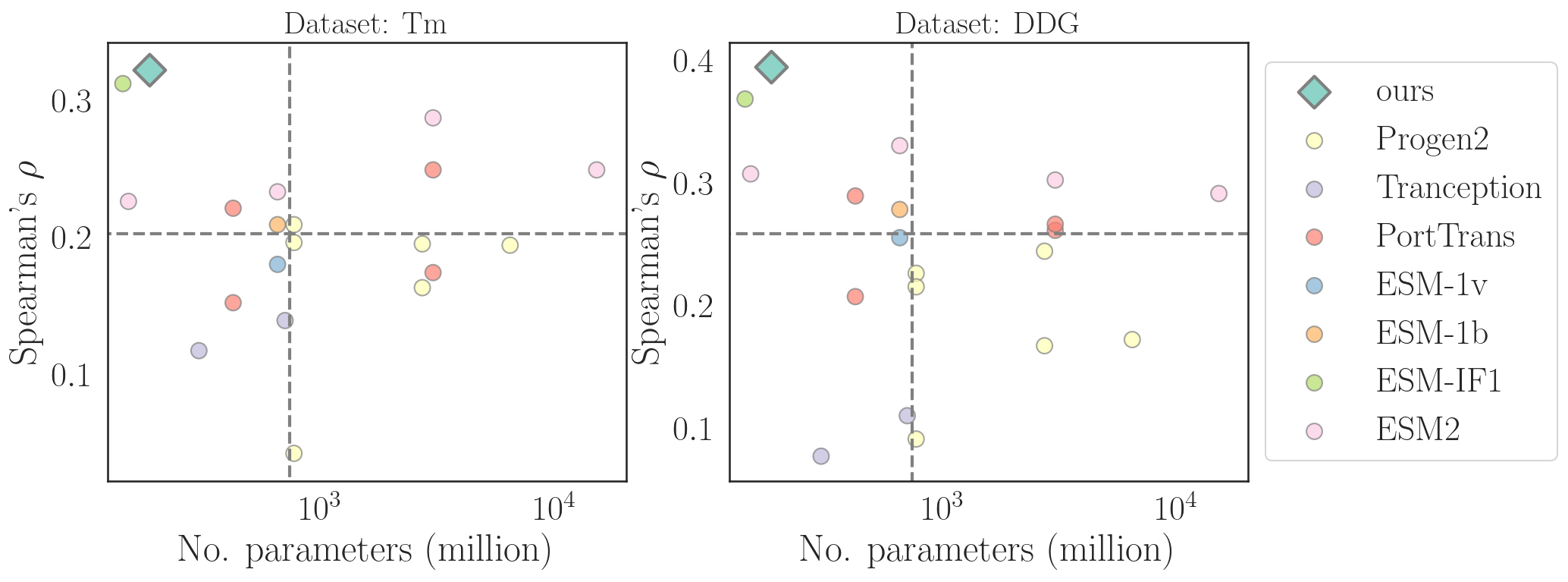}
    \caption{Number of parameters versus Spearman's $\rho$ correlation on \textbf{DTm} and \textbf{DDG}.}
    \label{fig:params}
\end{figure}

\subsection{Variant Effect Prediction}
Our model has demonstrated exceptional predictive performance compared to other SOTA models in forecasting the stability of protein mutation sequences in both \textbf{DTm} and \textbf{DDG}. \plg learns residue graphs with $k=20$ and deploys $1,280$ hidden neurons in each EGNN layer. Table~\ref{tab:TM_DMS} evaluates $100$ protein assays using TPR at $5\%$, $25\%$, and $50\%$, wherein \plg consistently outperforms competitors of varying model sizes. To further examine how our model efficiently achieves top performance relative to other large models, Figure~\ref{fig:params} visualizes Spearman’s correlation from predictions of pre-trained models at different model scales. Our model occupies the most desirable upper-left corner spot, where it reaches top-rank correlation with minimal computational cost, or equivalently, the smallest number of parameters to learn.

In addition to single-site predictions, we also test \plg's performance on deep mutations using $86$ protein assays in \textbf{ProteinGym} and compare its ranking with $27$ baselines. For the Spearman's correlation scores reported in Table~\ref{tab:ProteinGYM_DMS}, we reproduce \textsc{ESM} series (including \textsc{MSA-Transformer}) and \textsc{ProtTrans}, and retrieve scores for the remaining methods from \textbf{ProteinGym}'s repository \footnote{\url{https://github.com/OATML-Markslab/ProteinGym}}. Our method consistently predicts the most aligned ranks, regardless of mutational depth or the predicted taxon. Notably, we include 6 additional MSA-based models, which require fewer parameters but significantly longer inference times due to the need to query and process supplementary information in MSA for the target protein. Consequently, MSA-based methods achieve the second-best overall performance on \textbf{ProteinGym}, following closely behind \plg.

% Tab2 Results on ProteinGym benchmark which contains 86 proteins. Note that we exclude the longest protein of over $3,000$ AAs that fails to be folded by \textsc{AlphaFold2}.

\begin{table}[!t]
\caption{Variant Effect Prediction on \textbf{ProteinGym}.}
\label{tab:ProteinGYM_DMS}
\vspace{-5mm}
\begin{center}
\resizebox{\textwidth}{!}{
    \begin{tabular}{ccccccccccc}
    \toprule
    & \multirow{2}{*}{\textbf{Model}} & \multirow{2}{*}{\textbf{Version}} & \multirow{2}{*}{\textbf{\# Params} } & \multicolumn{3}{c}{\textbf{$\rho$ (by depth) $\uparrow$}} & \multicolumn{4}{c}{\textbf{$\rho$ (by taxon) $\uparrow$}} \\\cmidrule(lr){5-7}\cmidrule(lr){8-11}
    &  &  & (million) & Single & Double & All & Prokaryote & Human & Eukaryote & Virus \\
    \midrule
    \multirow{6}{*}{\rotatebox[origin=c]{90}{MSA}} 
    & \textsc{SiteIndep} & - & - & 0.378 & 0.322 & 0.378 & 0.343 & 0.375 & 0.401 & \textbf{0.406} \\
    & \textsc{EVmutation} & - & - & \textcolor{violet}{\textbf{0.423}} & \textcolor{red}{\textbf{0.401}} & \textcolor{violet}{\textbf{0.423}} & 0.499 & 0.396 & 0.429 & 0.381 \\
    & \textsc{Wavenet} & - & - & 0.399 & 0.344 & 0.400 & 0.492 & 0.373 & 0.442 & 0.321 \\
    & \textsc{DeepSequence} & - & - & \textbf{0.411} & 0.357 & \textbf{0.415} & 0.497 & 0.396 & \textcolor{violet}{\textbf{0.461}} & 0.332 \\ \cmidrule(lr){2-11}
     & \multirow{2}{*}{\textsc{MSA-Transfomer}} 
     & msa1 & 100 & 0.310 & 0.232 & 0.308 & 0.292 & 0.302 & 0.392 & 0.278 \\
    &  & msa1b & 100 & 0.291 & 0.275 & 0.290 & 0.268 & 0.282 & 0.365 & 0.279 \\
    \midrule
    \multirow{18}{*}{\rotatebox[origin=c]{90}{non-MSA}} & \multirow{4}{*}{RITA} & small & 85 & 0.324 & 0.211 & 0.329 & 0.311 & 0.314 & 0.330 & 0.372 \\
    &  & medium & 300 & 0.372 & 0.237 & 0.377 & 0.356 & 0.370 & 0.399 & 0.398 \\
    &  & large & 680 & 0.372 & 0.227 & 0.383 & 0.353 & 0.380 & 0.404 & 0.405 \\
    &  & xlarge & 1,200 & 0.385 & 0.234 & 0.389 & 0.405 & 0.364 & 0.393 & \textcolor{red}{\textbf{0.407}} \\\cmidrule(lr){2-11}
    & \multirow{5}{*}{\textsc{ProGen2}} & small & 151 & 0.346 & 0.249 & 0.352 & 0.364 & 0.376 & 0.396 & 0.273 \\
    &  & medium & 764 & 0.394 & 0.274 & 0.395 & 0.434 & 0.393 & 0.411 & 0.346 \\
    &  & base & 764 & 0.389 & 0.323 & 0.394 & 0.426 & 0.396 & 0.427 & 0.335 \\
    &  & large & 2,700 & 0.396 & 0.333 & 0.396 & 0.431 & 0.396 & 0.436 & 0.336 \\
    &  & xlarge & 6,400 & 0.404 & 0.358 & 0.404 & 0.480 & 0.349 & 0.452 & 0.383 \\ \cmidrule(lr){2-11}
    & \multirow{4}{*}{\textsc{PortTrans}} & bert & 420 & 0.339 & 0.279 & 0.336 & 0.403 & 0.300 & 0.345 & 0.317 \\
    &  & bert\_bfd & 420 & 0.311 & 0.336 & 0.308 & 0.471 & 0.328 & 0.338 & 0.087 \\
    &  & t5\_xl\_uniref50 & 3,000 & 0.384 & 0.284 & 0.378 & 0.485 & 0.375 & 0.369 & 0.277 \\
    &  & t5\_xl\_bfd & 3,000 & 0.355 & 0.356 & 0.351 & 0.490 & 0.399 & 0.349 & 0.131 \\
    \cmidrule(lr){2-11}
    & \textsc{Tranception} & large & 700 & 0.399 & \textcolor{violet}{\textbf{0.398}} & 0.406 & 0.447 & 0.369 & 0.426 & \textcolor{violet}{\textbf{0.407}} \\\cmidrule(lr){2-11}
    & ESM-1v & - & 650 & 0.376 & 0.290 & 0.372 & 0.496 & 0.409 & 0.398 & 0.233 \\\cmidrule(lr){2-11}
    & \textsc{ESM-1b} & - & 650 & 0.371 & 0.325 & 0.366 & \textbf{0.507} & 0.416 & 0.360 & 0.150 \\ \cmidrule(lr){2-11}
    & \textsc{ESM-if1} & - & 142 & 0.359 & 0.279 & 0.368 & 0.445 & 0.358 & 0.339 & 0.322 \\\cmidrule(lr){2-11}
    & \multirow{4}{*}{\textsc{ESM-2}} & t30 & 150 & 0.345 & 0.296 & 0.344 & 0.437 & \textbf{0.419} & 0.401 & 0.045 \\
    &  & t33 & 650 & 0.392 & 0.317 & 0.389 & \textcolor{violet}{\textbf{0.515}} & \textcolor{red}{\textbf{0.433}} & \textbf{0.454} & 0.155 \\
    &  & t36 & 3,000 & 0.384 & 0.261 & 0.383 & 0.495 & 0.419 & 0.429 & 0.195 \\
    &  & t48 & 15,000 & 0.394 & 0.313 & 0.391 & 0.457 & 0.402 & 0.442 & 0.251 \\ \cmidrule(lr){2-11}
    & \plg & k20\_h512 & 148 & \textcolor{red}{\textbf{0.424}} & \textbf{0.395} & \textcolor{red}{\textbf{0.426}} & \textcolor{red}{\textbf{0.516}} & \textcolor{violet}{\textbf{0.425}} & \textcolor{red}{\textbf{0.480}} & 0.297\\
    \bottomrule\\[-2.5mm]
    \multicolumn{11}{l}{$\dagger$ The top three are highlighted by \textbf{\textcolor{red}{First}}, \textbf{\textcolor{violet}{Second}}, \textbf{Third}.}
    \end{tabular}
}
\end{center}
\end{table}

\subsection{Ablation Study}
\label{sec:experiment_ablation}
This section evaluates the prediction performance of \textbf{ProteinGym} based on Spearman's correlation on different modular designs of \plg. The results are visualized in Figure~\ref{fig:ablationStudy} with additional details supplemented in Appendix~\ref{sec:app:ablation}. In this section, we mix the inference results for each primary criterion with diverse secondary arguments. For instance, in the top orange box of Figure~\ref{fig:ablationStudy}(a), we report all ablation results that utilize $6$ EGNN layers for graph convolution, regardless of the different scales of \textsc{ESM-2} or the definitions of node attributes. For all modules investigated in this section, we separately discuss their influence on predicting mutational effects when modifying a single site or an arbitrary number of sites. These two cases are marked respectively as `single' and `all' on the y-axis. 

\begin{figure}[!t]
    \centering
    \includegraphics[width=\textwidth]{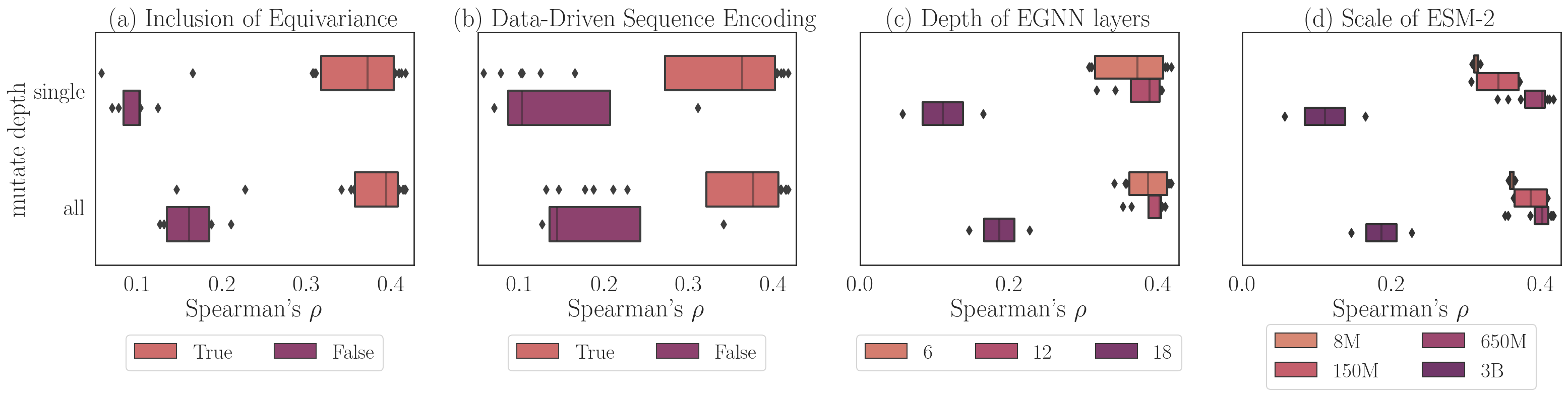}
    \caption{Ablation Study on \textbf{ProteinGym}, evaluated by Spearman's correlation on single-site and deep mutations.}
    \vspace{-5mm}
    \label{fig:ablationStudy}
\end{figure}

\paragraph{Inclusion of Roto-Translation Equivariance} 
We assess the effect of incorporating rotation and translation equivariance during protein geometric and topological encoding. Three types of graph convolutions are compared, including GCN \cite{Kipf2017semi}, GAT \cite{velivckovic2017graph}, and EGNN \cite{satorras2021n}. The first two are classic non-equivariant graph convolutional methods, while the last one, which we apply in the main algorithm, preserves roto-translation equivariance. We fix the number of EGNN layers to $6$ and examine the performance of the other two methods with either $4$ or $6$ layers. We find that integrating equivariance when embedding protein geometry significantly improves prediction performance.

\paragraph{Sequence Encoding}
We next investigate the benefits of defining data-driven node attributes for protein representation learning. We compare the performance of models trained on two sets of graph inputs: the first set defines its AA node attributes through trained \textsc{ESM2} \cite{lin2023esm2},  while the second set uses one-hot encoded AA types for each node. A clear advantage of using hidden representations by prefix models over hardcoded attributes is evident from the results presented in Figure~\ref{fig:ablationStudy}(b).

\paragraph{Depth of EGNN}
Although graph neural networks can extract topological information from geometric inputs, it is vital to select an appropriate number of layers for the module to deliver the most expressive node representation without encountering the oversmoothing problem. We investigate a wide range of choices for the EGNN layers among $\{6,12,18\}$. As reported in Figure~\ref{fig:ablationStudy}(c), embedding graph topology with deeper networks does not lead to performance improvements. A moderate choice of $6$ EGNN layers is sufficient for our learning task. 

\paragraph{Scale of ESM}
We also evaluate our models on different choices of language embedding dimensions to study the trade-off between the computational cost and input richness. 
Various scales of prefix models, including $\{8, 150, 650, 3000\}$ millions of parameters, have been applied to produce different sequential embeddings with $\{320,640,1280,2560\}$ dimensions, respectively. Figure~\ref{fig:ablationStudy}(d) reveals a clear preference for \textsc{ESM-2}-t33, which employs $650$ million parameters to achieve optimal model performance with the best stability. Notably, a higher dimension and richer semantic expression do not always yield better performance. In fact, performance degradation is observed when using the t36 version of the prefix model with 3 billion parameters.
% In order to study the effect of the choice of the initial point on the performance of graph network and trade off the computation cost, we select different scales of ESM as the prefix models. The scales are set to \textbf{(8M, 150M, 650M and 3B)}. It seems there is a bottleneck between the size and rich representation

\section{Related Work}
% \bxc{consider connecting the three parts of related work in the opening of this section. For instance: the first two subsections review the general embedding rules for 2D and 3D representations of proteins, \ie sequence and structure}

% \bx{+ref: DeepFRI: Gligorijević, V., Renfrew, P. D., Kosciolek, T., Leman, J. K., Berenberg, D., Vatanen, T., Chandler, C., Taylor, B. C., Fisk, I. M., Vlamakis, H., Xavier, R. J., Knight, R., Cho, K., & Bonneau, R. (2021). Structure-based protein function prediction using graph convolutional networks. Nature Communications, 12(1), 1-14. https://doi.org/10.1038/s41467-021-23303-9}

\paragraph{Protein Primary Structure Embedding}
Self-supervised protein language models play the predominant role in the training of large quantities of protein sequences for protein representation learning. These methodologies draw parallels between protein sequences and natural language, encoding amino acid tokens using the \textsc{Transformer} model \cite{vaswani2017attention} to extract pairwise relationships among tokens. These methods typically pre-train on extensive protein sequence databases to autoregressively recover protein sequences \cite{madani2023progen,notin2022tranception}.
Alternatively, masked language modeling objectives develop attention patterns that correspond to the residue-residue contact map of the protein \cite{lin2023esm2,meier2021esm1v,rao2021transformer,rives2021esm1b,vig2021bertology}. Other methods start from a multiple sequence alignment, summarizing the evolutionary patterns in target proteins \cite{frazer2021disease,rao2021msa,riesselman2018deepsequence}.
Both aligned and non-aligned methods result in a strong capacity for discovering the hidden protein space, but this often comes at the expense of excessive training input or the use of substantial learning resources. This trade-off underlines the need for efficient and cost-effective approaches in self-supervised protein modeling.

\paragraph{Protein Tertiary Structure Embedding}
Protein structures directly dictate protein functions and are essential to \textit{de novo} protein design, which is a critical challenge in bioengineering. The remarkable success of accurate protein folding by \textsc{AlphaFold2} \cite{jumper2021alphafold} and the subsequent enrichment of the structure-aware protein repository have motivated a series of research initiatives focused on learning protein geometry. Recent efforts have been made to encode geometric information of proteins \cite{gligorijevic2021fpgcn,jing2020gvp,zhang2022gsp} for topology-sensitive tasks such as molecule binding \cite{jin2021iterative, kong2023conditional, myung2022csm}, protein interface analysis \cite{mahbub2022egret, reau2023deeprank}, and protein properties prediction \cite{zhang2022ontoprotein}. 

% \paragraph{Protein Sequence Recovery}
% A common learning task for self-supervised learning models: recover the full protein sequence.

\paragraph{Variant Effect Prediction}
Variant effect predictions quantify the fitness of a mutant protein in comparison to its wild-type counterpart. For well-studied proteins, it is feasible to fit a supervised model on the hidden representation of amino acid local environments to infer fitness scores \cite{frazer2021eve,lu2022machine,marquet2022vespa,zhou2023accurate}. However, in many cases, labeled data is either scarce or inaccessible. To overcome this, zero-shot methods have been developed to infer the fitness of a mutation from the evolutionary landscape of the original protein using sequence alignments \cite{hopf2017evmutation, riesselman2018deepsequence, shin2021wavenet}. Alternatively, hybrid models \cite{notin2022tranception, rao2021msa} utilize retrieval or attention mechanisms to extract patterns from Multiple Sequence Alignments (MSAs) in conjunction with protein language or structure models.

% Alignment-based models \cite{riesselman2018deepsequence, shin2021wavenet, hopf2017evmutation, frazer2021eve} extract information from MSAs across thousands of protein families, and hybrid models \cite{rao2021msa, notin2022tranception} make use of retrieval or attention machine to extract patterns from MSAs combined with protein language or structure models. Another series of work use the embeddings  extracted from large-scale PLMs with a supervised \cite{marquet2022vespa} or unsupervised \cite{meier2021esm1v} framework.

\section{Conclusion and Discussion}
\label{sec:discussion}
The development of dependable computational methodologies for protein engineering is a crucial facet of \textit{in silico} protein design. Accurately assessing the fitness of protein mutants not only supports cost-effective experimental validations but also guides the modification of proteins to enhance existing or introduce new functions. Most recent deep learning solutions employ a common strategy that involves establishing a hidden protein representation and masking potential mutation sites for amino acid generation. Previous research has primarily focused on extracting protein representations from either their sequential or structural modalities, with many treating the prediction of mutational effects merely as a secondary task following inverse folding or \textit{de novo} protein design. These approaches often overlook the importance of comprehending various levels of protein structures that are critical for determining protein function. Furthermore, they seldom implement model designs tailored specifically for mutation tasks. In this work, we introduce \plg, a denoising framework that effectively cascades protein primary and tertiary structure embedding for the specific task of predicting mutational effects. This framework first employs a prefix protein language model to decode sequence representation and identify residue-wise intercommunications. This is subsequently enhanced by a roto-translation equivariant graph neural network, which encodes geometric representations for amino acid microenvironments. We have extensively validated the efficacy of \plg across various protein function assays and taxa, including two thermal stability databases that were prepared by ourselves. Our approach consistently demonstrates substantial promise for protein engineering applications, particularly in facilitating the design of mutation sequences with improved thermal stability.

\paragraph{Broader Impact} 
The intersection of deep learning and structural biology, as showcased in this study, has the potential to transform our approach to protein engineering challenges, paving the way for sustainable and efficient solutions. Algorithms, such as \plg, are primarily designed to enhance enzymes to support initiatives in drug discovery, biofuel production, and other relevant industries. However, given the pervasive presence of proteins across numerous scenarios and organisms, it is feasible that these methods could be employed to modify dangerous species, such as harmful viruses. Therefore, it is crucial to regulate the use of these deep learning methods, akin to the oversight required for any other powerful tools. Interestingly, our model demonstrates suboptimal performance when applied to such categories of proteins (refer to Table~\ref{tab:ProteinGYM_DMS}), suggesting an inherent limitation in its potential misuse. 

% \plg is designed for mutation effect evaluation but may be used for the design of more sophisticated virus which is harmful to our society. However, luckily our model does not perform outstandingly on this taxon. 

\paragraph{Limitation}
The consumption of training resources for AI-driven protein engineering techniques has surged considerably nowadays. For instance, \textsc{ESM-IF1}, which is another geometric model that utilizes structural information of proteins, necessitates months of processing time and hundreds of machines to integrate sequence and topological data. Owing to these computational cost constraints, our approach does not train on such an extensive corpus from scratch. Instead, we harness publicly-available language models to extract hidden representations for amino acids.
Nevertheless, training and inference in such an integrated model require geometric information from proteins in addition to sequential data. The current data repositories are rich with protein structures experimentally solved by biologists and supplemented by high-quality protein geometries from contemporary techniques such as \textsc{AlphaFold2}, and they are adequate for training our model. However, it's plausible that a revised protein could have an excessively long sequence that lacks a crystallized structure and cannot be folded by computational tools. An example of such a limitation is evident in our experiment, where a protein with an extended length was removed from the \textbf{ProteinGym} database.

\bibliography{main.bbl}
\bibliographystyle{plain}

%%%%%%%%%%%%%%%%%%%%%%%%%%%%%%%%%%%%%%%%%%%%%%%%%%%%%%%%%%%%
% \input{checklist}

\newpage
\appendix
\input{appendix}

\end{document}

%% file: math_commands.tex
\usepackage{amsmath,amsfonts,bm}

\def\eqref#1{equation~\ref{#1}}
\def\1{\bm{1}}

\def\vm{{\bm{m}}}

\def\vv{{\bm{v}}}
\def\vw{{\bm{w}}}
\def\vx{{\bm{x}}}
\def\vy{{\bm{y}}}

\def\mV{{\bm{V}}}
\def\mW{{\bm{W}}}
\def\mX{{\bm{X}}}
\def\mY{{\bm{Y}}}

\DeclareMathAlphabet{\mathsfit}{\encodingdefault}{\sfdefault}{m}{sl}
\SetMathAlphabet{\mathsfit}{bold}{\encodingdefault}{\sfdefault}{bx}{n}

\def\gE{{\mathcal{E}}}

\def\gG{{\mathcal{G}}}

\def\gN{{\mathcal{N}}}

\def\gT{{\mathcal{T}}}

\def\gV{{\mathcal{V}}}

\newcommand{\R}{\mathbb{R}}

%% file: appendix.tex
\section{Dataset Description}
\label{sec:app:Tm_DDG}
\subsection{New Benchmarks: DTm and DDG}
\begin{figure}[!ht]
    \centering
    \includegraphics[width=\textwidth]{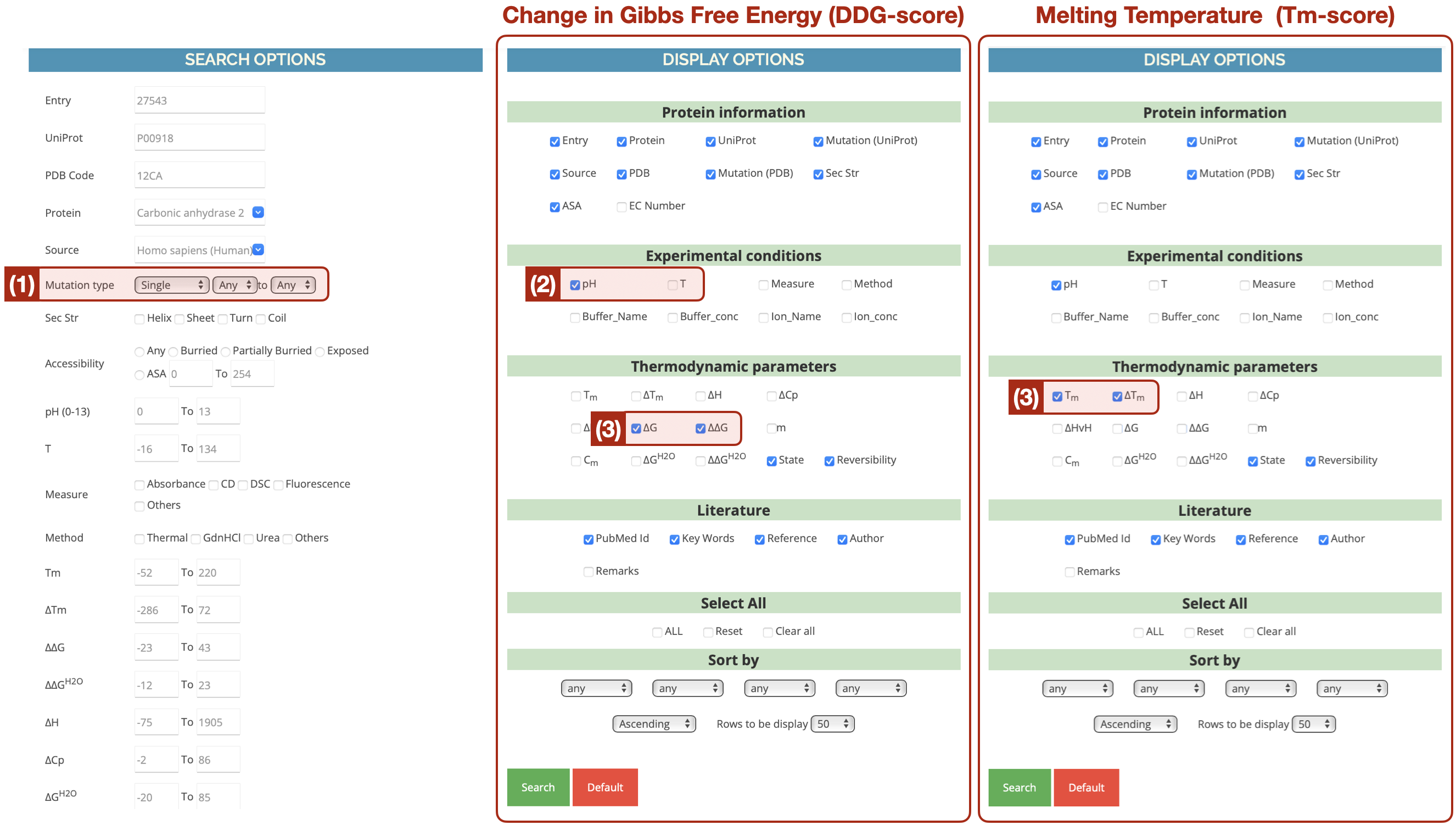}
    \caption{Specifications of mutational records sourced from \textbf{ProThermDB}. We (1) limit the number of mutational sites to ``single" and (2) require the experimental conditions to be defined by the pH level. Regarding thermodynamic parameters, we incorporate (3) changes ($\Delta$G) and the change of changes ($\Delta\Delta$G) in Gibbs Free Energy, and melting temperature (Tm) and changes of Tm ($\Delta$Tm) for \textbf{DDG} and \textbf{DTm}, respectively.  }
    \label{fig:proThermDB}
\end{figure}

We have established two novel benchmarks, namely \textbf{DTm} and \textbf{DDG}, to assess model accuracy in predicting the stability of proteins that have single-site mutations. The proteins included in both benchmarks have been sourced from \textbf{ProThermDB} \cite{nikam2021prothermdb}. Given the profound influence experimental conditions exert on protein stability, we have made it a point to denote the pH (potential of hydrogen) environment and have assigned each protein-environment pairing a name that follows the format ``\textit{PDB ID-pH level}" (refer to Figure~\ref{fig:proThermDB}). For instance, ``\textit{O00095-8.0}" in \textsc{DDG} signifies that the mutational records were conducted and evaluated under a pH of $8.0$ for protein \texttt{O00095}. 

\begin{figure}[h]
    \centering
    \includegraphics[width=0.8\textwidth]{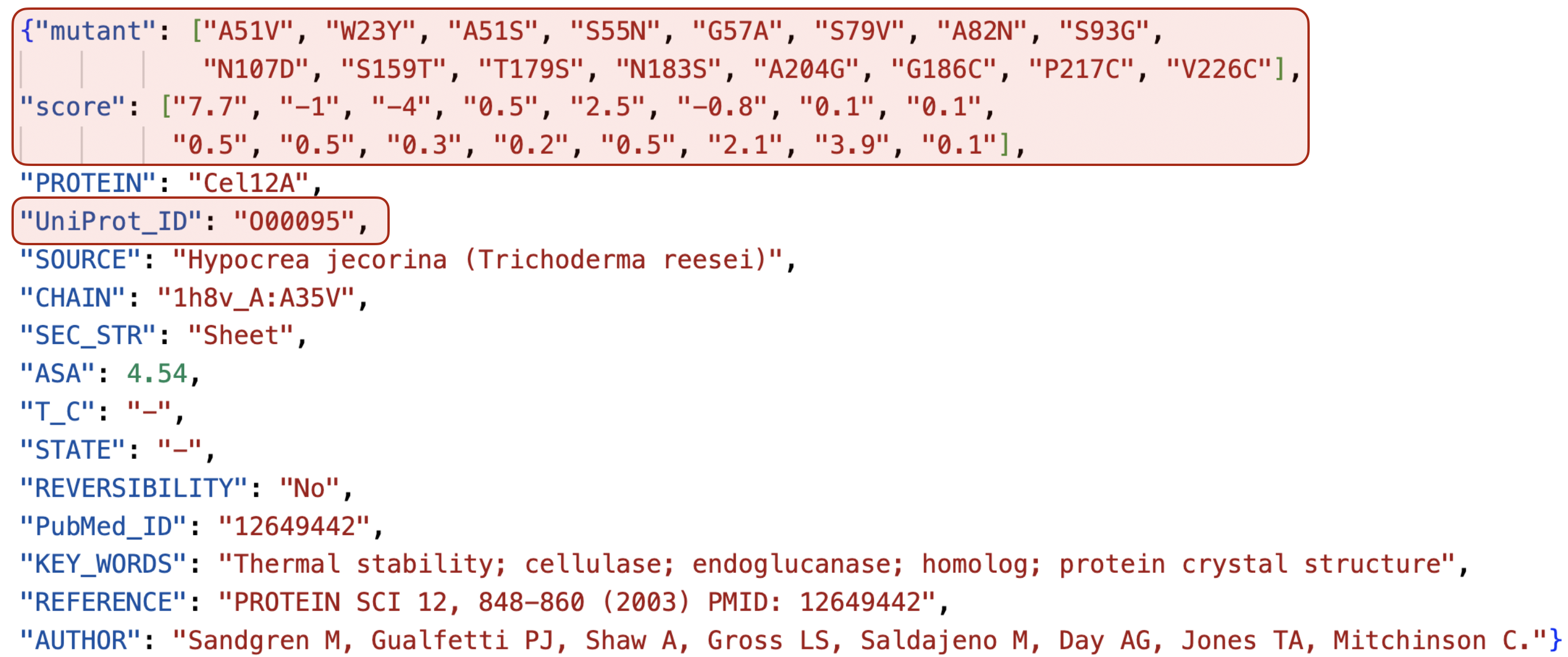}
    \caption{An example source record of mutational assay.}
    \label{fig:o00095}
\end{figure}

A sample of the acquired ``protein-environment" source records is presented in Figure~\ref{fig:o00095}. We further process the attributes ``mutant", ``score", and ``UniProt\_ID". Initially, we curated all sets in \textbf{ProThermDB} containing at least $10$ records to achieve statistically significant correlation and TPR evaluations. Next, we manually scrutinized the mutation points relative to their raw sequences and removed records with continuous mutations. In other words, we exclude mutations upon mutations, concentrating instead on the single-site mutations compared to wild-type proteins. Furthermore, we employ UniProt ID to pair protein sequences with folded structures predicted by \textsc{AlphaFold2} \cite{jumper2021alphafold} from \url{https://alphafold.ebi.ac.uk}. We abstain from querying the protein structure by their PDB ID to prevent dealing with partial or misaligned protein sequences due to incompleteness in wet experimental procedures. In total, \textbf{DTm} consists of $60$ such protein-environment pairs, and \textbf{DDG} encompasses $30$. %The designation and the number of mutations are detailed for each pair in Figure~\ref{fig:distBenchmark}.

% \begin{figure}[!t]
%     \centering
%     \includegraphics[width=0.8\textwidth]{}
%     \caption{Frequencies of ``protein-pH" combinations in \textbf{DTm} and \textbf{DDG}.}
%     \label{fig:distBenchmark}
% \end{figure}

\subsection{From \textsc{Tranception}: ProteinGym}
We also validate our model's ability to predict deep mutational effects using \textbf{ProteinGym} \cite{notin2022tranception}. It is currently the most extensive protein substitution benchmark, comprising roughly $1.5M$ missense variants across $87$ DMS assays. These DMS assays cover a broad spectrum of functional properties (\eg, thermostability, ligand binding, aggregation, viral replication, drug resistance) and span diverse protein families (\eg, kinases, ion channel proteins, g-protein coupled receptors, polymerases, transcription factors, tumor suppressors) across different taxa (\eg, humans, other eukaryotes, prokaryotes, viruses). In this study, we evaluated $86$ variants from the full set. We excluded one protein, namely \texttt{A0A140D2T1\_ZIKV\_Sourisseau\_growth\_2019}, due to its failure to be folded by \textsc{AlphaFold2}.

\section{Noise Addition Strategies}
\label{sec:app:noise}
In the primary algorithm, noise sampling is utilized at each epoch on the node attributes to emulate random mutations in nature. This introduction of noise directly affects the node attribute input to the graphs. Alongside the ``mutate-then-recode" method we implemented in the main algorithm, we examined four additional strategies to perturb the input data during training. The construction of these strategies is detailed below, and the corresponding model performance is reported in Table~\ref{tab:noiseAddition}, which returns Spearman's correlation on \textbf{ProteinGym}.

\begin{table}[!ht]
\caption{Performance comparison on different noise addition strategies.}
\label{tab:noiseAddition}
\begin{center}
% \resizebox{\textwidth}{!}{
    \begin{tabular}{cccccc}
    \toprule
    strategy & \textbf{Mean} & \textbf{Sliding Window} & \textbf{Gaussian Noise} & \textbf{Mask} & \textbf{Mutate\&Recode} \\
    \midrule
    $\rho$ & 0.245 & 0.215 & 0.229 & 0.396 & \textbf{0.426} \\
    \bottomrule\\[-2.5mm]
    \end{tabular}
% }
\end{center}
\end{table}

\paragraph{Mean}
Suppose the encoded sequential representation of a node is predominantly determined by its amino acid (AA) type. In essence, the protein sequence encoder will return similar embeddings for nodes of the same AA type, albeit at different positions within a protein. With this premise, the first method replaces the node embedding for perturbed nodes with the average representations from the same AA types. For example, a random node $\vv_i$ in a protein is of type L. If it is altered in the current epoch, $\tilde{\vv}_i$ is designated as the average sequential embedding of all other nodes of type L.

\paragraph{Sliding Window}
The presumption in the previous method neither aligns with the algorithmic design nor biological heuristics. Self-attention discerns the interaction of the central token with its neighbors across the entire document (protein), and AAs, inclusive of their types and properties, are thought to be closely aligned with its local environment. Thus, averaging embeddings of AAs from varying positions is likely to forfeit positional information of the modified AA. Consequently, the second method designs a sliding window along the protein sequence for perturbing node noise. By setting the window size to $k$, a corrupted node will update its representation by averaging the node representation of the nearest $k$ neighbor AAs along the AA chain.

\paragraph{Gaussian Noise}
The subsequent method regards node embeddings of AA as hidden vectors and imposes white noise on the vector values. We define the mean and variance of the noise as 0 and 0.5, respectively, making the revised node representation $\tilde{\vv}=\vv+\gN(0,0.5)$. 

\paragraph{Mask}
Finally, we employ the masking technique prevalent in masked language modeling and substitute the perturbed AA token with a special \texttt{<mask>} token. The prefix language model will interpret it as a novel token and employ self-attention mechanisms to assign a new representation to it. We utilize the same hyperparameter settings as that of BERT \cite{devlin2018bert} and choose $15\%$ of tokens per iteration as potential influenced subjects. Specifically, $80\%$ of these subjects are replaced with \texttt{<mask>}, $10\%$ of them are replaced randomly (to other AA types), and the rest $10\%$ stay unchanged.

\section{Baseline Method Implementation Details}
\label{sec:app:baselineImplementation}
In our series of experiments, we compared a wide array of baseline methods across three benchmarks. All baseline methods were evaluated on two new datasets, namely \textbf{DTm} and \textbf{DDG}. We conducted these evaluations by independently reproducing the methods based on the algorithms or pre-trained models available in their respective official repositories. The names of the models along with pertinent details are provided in Table~\ref{tab:models}.

\begin{table}[!ht]
\caption{Details of Baseline Models.}
\label{tab:models}
\begin{center}
\resizebox{\textwidth}{!}{
    \begin{tabular}{lll}
    \toprule
    Model & Description & URL
    \\
    \midrule
    \textsc{DeepSequence} \cite{riesselman2018deepsequence} & \makecell[l]{a VAE-based model trained for every\\target protein with their MSAs} & \url{https://github.com/debbiemarkslab/DeepSequence} \\\midrule
    \textsc{Tranception} \cite{notin2022tranception} & \makecell[l]{an autoregressive model for variant\\effect prediction task with retrieve machine} & \url{https://github.com/OATML-Markslab/Tranception} \\\midrule
    \textsc{ESM-1b} \cite{rives2021esm1b} & \multirow{4}{*}{\makecell[l]{masked language model-based pre-train\\method with various pre-training dataset\\and positional embedding strategies}} & \multirow{7}{*}{\url{https://github.com/facebookresearch/esm}}\\
    \textsc{MSA-Transformer} \cite{rao2021msa} & & \\
    \textsc{ESM-1v} \cite{meier2021esm1v} & & \\
    \textsc{ESM2} \cite{lin2023esm2} && \\ \cmidrule(lr){1-2}
    \textsc{ESM-if1} \cite{hsu2022esmif} & \makecell[l]{an inverse folding method with both mask\\language modeling and Geometric Vector\\Perceptron (GVP)} & \\\midrule
    \textsc{RITA} \cite{hesslow2022rita} & \makecell[l]{a generative protein language model with\\billion-level parameters} & \url{https://github.com/lightonai/RITA}\\ \midrule
    \textsc{ProGen2} \cite{nijkamp2022progen2}  & \makecell[l]{a generative protein language model with\\billion-level parameters} & \url{https://github.com/salesforce/progen}\\\midrule
    \textsc{ProtTrans} \cite{ahmed2021prottrans} &  \makecell[l]{\textsc{Transformer}-based models trained on\\large protein sequence corpus} & \url{https://github.com/agemagician/ProtTrans} \\
    \bottomrule\\[-2.5mm]
    % \multicolumn{3}{l}{$\dagger$ The top three are highlighted by \textbf{\textcolor{red}{First}}, \textbf{\textcolor{violet}{Second}}, \textbf{Third}.}
    \end{tabular}
}
\end{center}
\end{table}

To ensure fair comparisons in the baseline models, we took two specific measures. First, the retrieval module in \textsc{Tranception} was removed when assessing its performance on the \textbf{ProteinGym} dataset. This step was taken as this model had been specifically tailored for this particular dataset. Second, the comprehensive version of \textsc{ESM-1v} averages model predictions from five different variants of diverse setups and parameters. However, as the rest of the baseline methods did not implement ensemble strategies, we tested \textsc{ESM-1v} using only its first variant.

For the rest methods we compared on the \textbf{ProteinGym} benchmark, the results are retrieved from \url{https://marks.hms.harvard.edu/proteingym/scores_all_models_proteingym_substitutions.zip} except for \textsc{MSA Transformer}, which is reproduced by ourselves.

\section{Varient Effect Prediction on Individual Proteins}
\begin{figure}[!ht]
    \centering
    \includegraphics[width=0.8\textwidth]{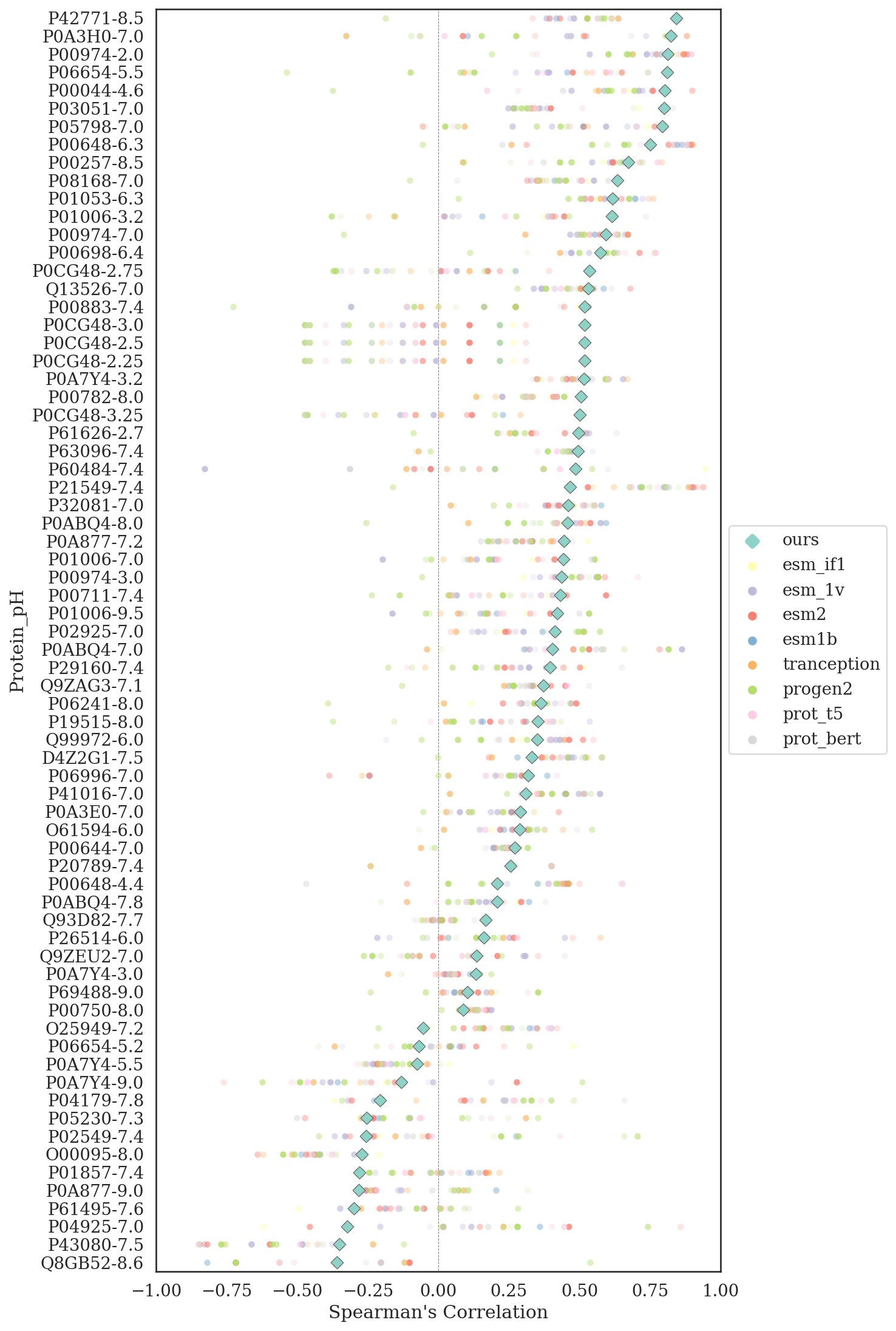}
    \caption{Protein-wise Spearman's correlation for \textbf{DTm}}
    \label{fig:Tm_corr}
\end{figure}

\begin{figure}[!ht]
    \centering
    \includegraphics[width=0.8\textwidth]{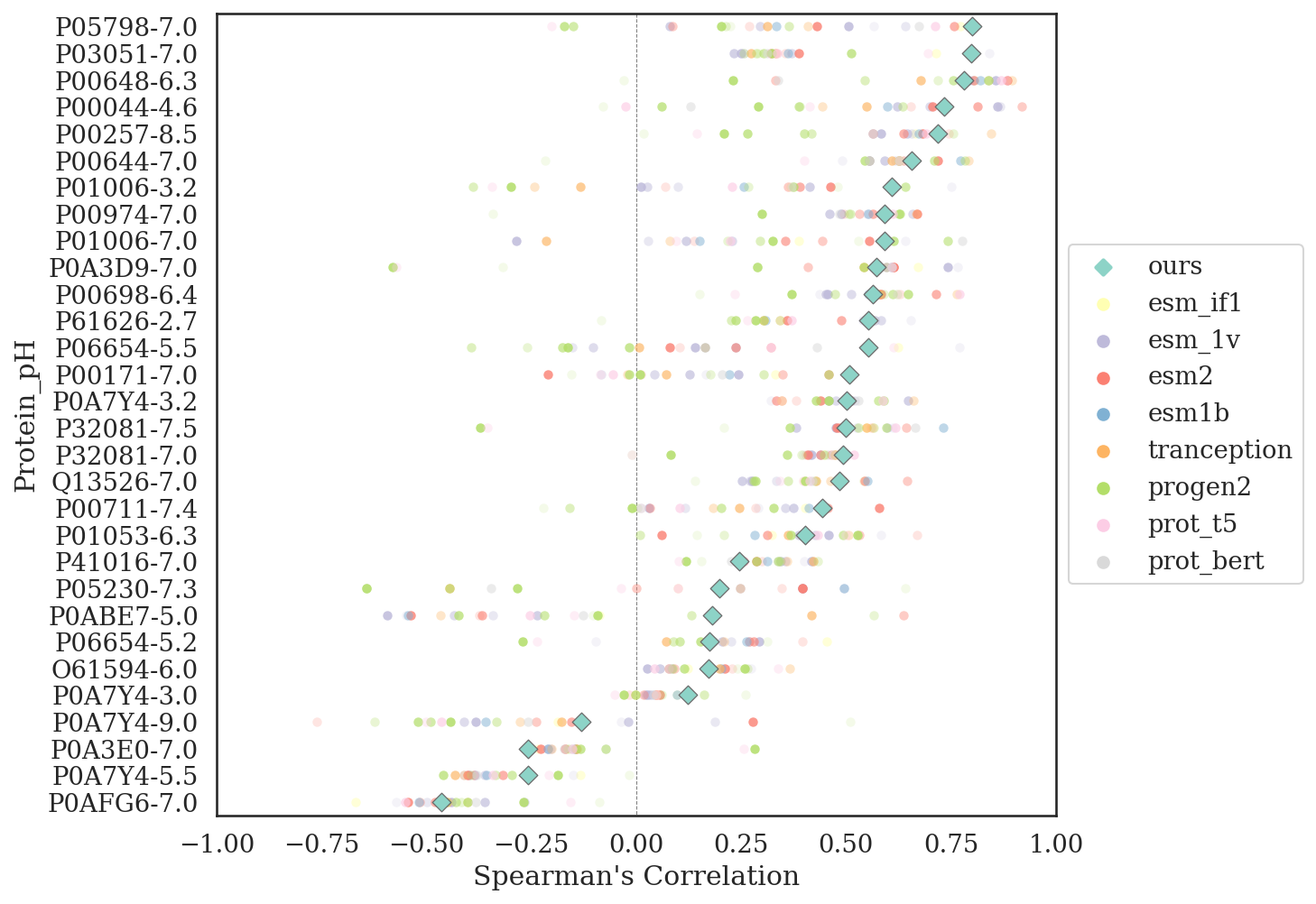}
    \caption{Protein-wise Spearman's correlation for \textbf{DDG}}
    \label{fig:ddg_corr}
\end{figure}

\begin{figure}[!ht]
    \centering
    \includegraphics[width=\textwidth]{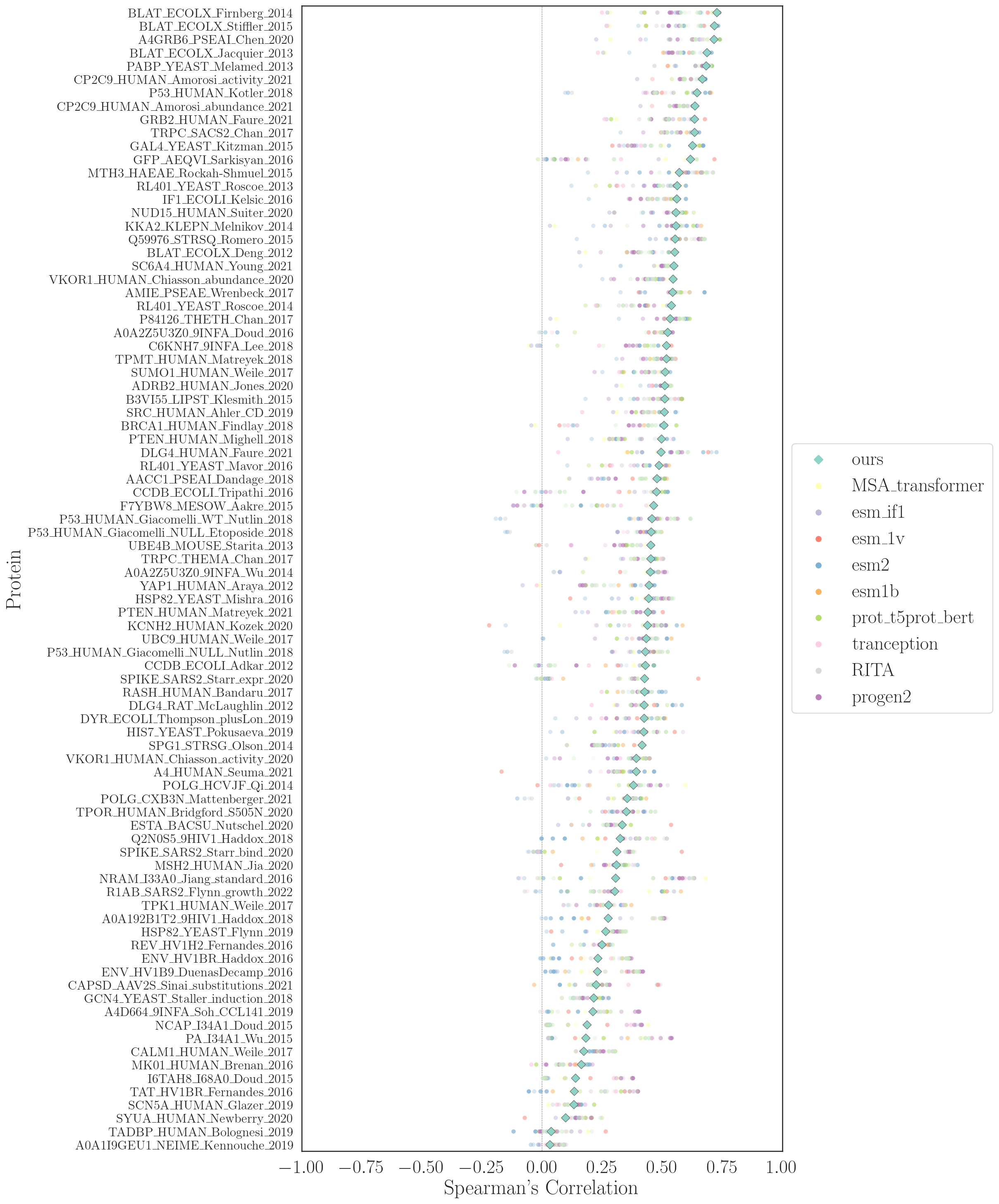}
    \caption{Protein-wise Spearman's correlation for \textbf{ProteinGym}}
    \label{fig:pg_corr}
\end{figure}

Figures~\ref{fig:Tm_corr} through~\ref{fig:pg_corr} illustrate the correlation performance for individual proteins within the \textbf{DTm}, \textbf{DDG}, and \textbf{ProteinGym} benchmarks. Results are color-coded according to their model types, with varying degrees of transparency indicating different versions of the same baseline method. Across the majority of proteins and protein-environment combinations, our developed \plg consistently outperforms competitor models. For \textbf{ProteinGym}, we have intentionally excluded alignment-based methods (including \textsc{SiteIndep}, \textsc{EVmutation}, \textsc{Wavenet}, and \textsc{DeepSequence}) from our comparisons, focusing instead on assessing the performance of general-purpose pre-trained models.

\begin{figure}[!ht]
    \centering
    \includegraphics[width=\textwidth]{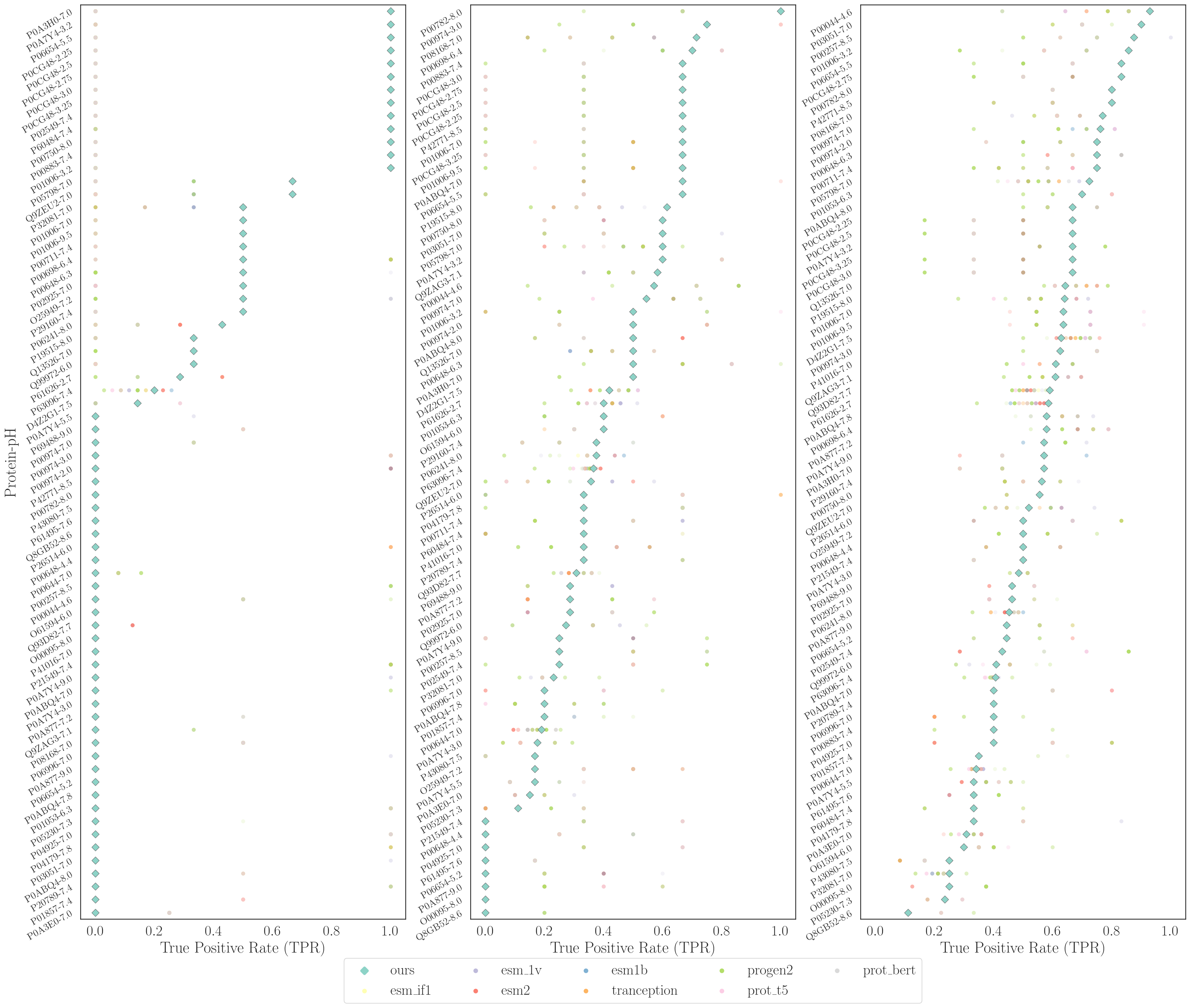}
    \caption{TPR on \textbf{DTm} with top $5\%$ (left), $25\%$ (middle), and $50\%$ (right) samples be identified positive.}
    \label{fig:Tm_tpr}
\end{figure}

\begin{figure}[!ht]
    \centering
    \includegraphics[width=\textwidth]{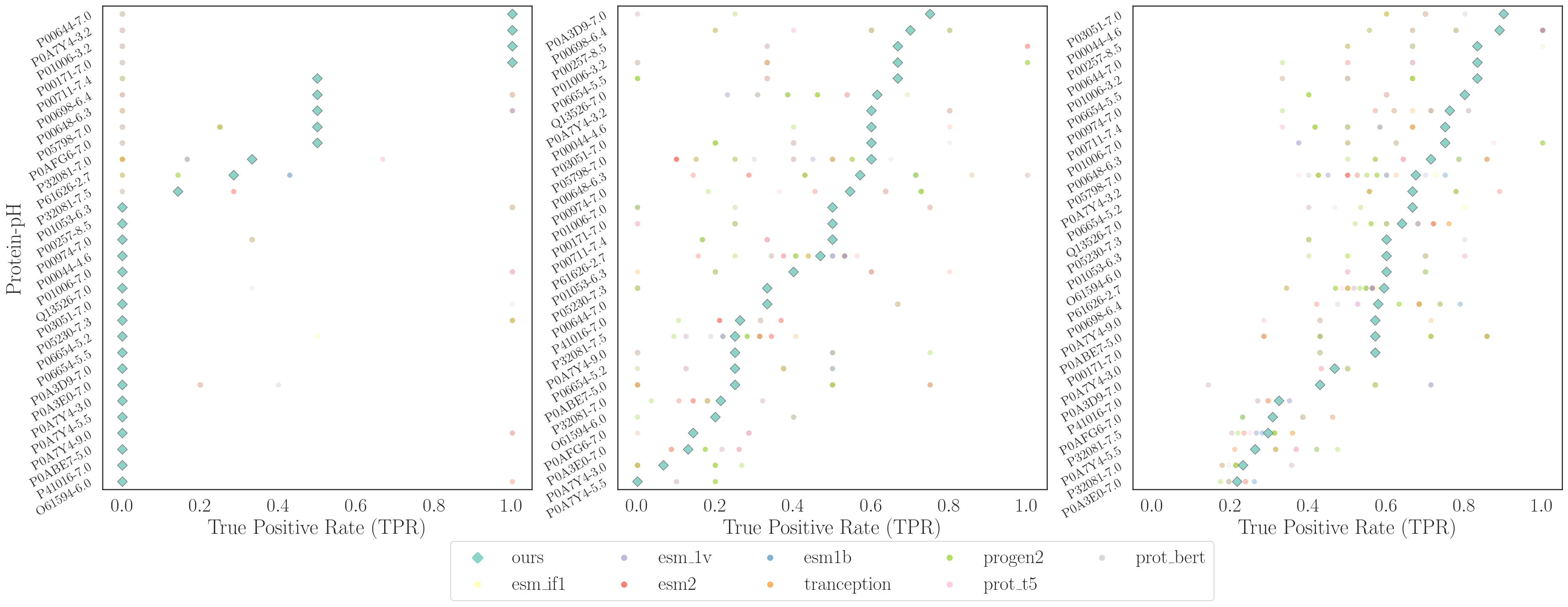}
    \caption{TPR on \textbf{DDG} with top $5\%$ (left), $25\%$ (middle), and $50\%$ (right) samples be identified positive.}
    \label{fig:ddg_tpr}
\end{figure}

\begin{figure}[!ht]
    \centering
    \includegraphics[width=\textwidth]{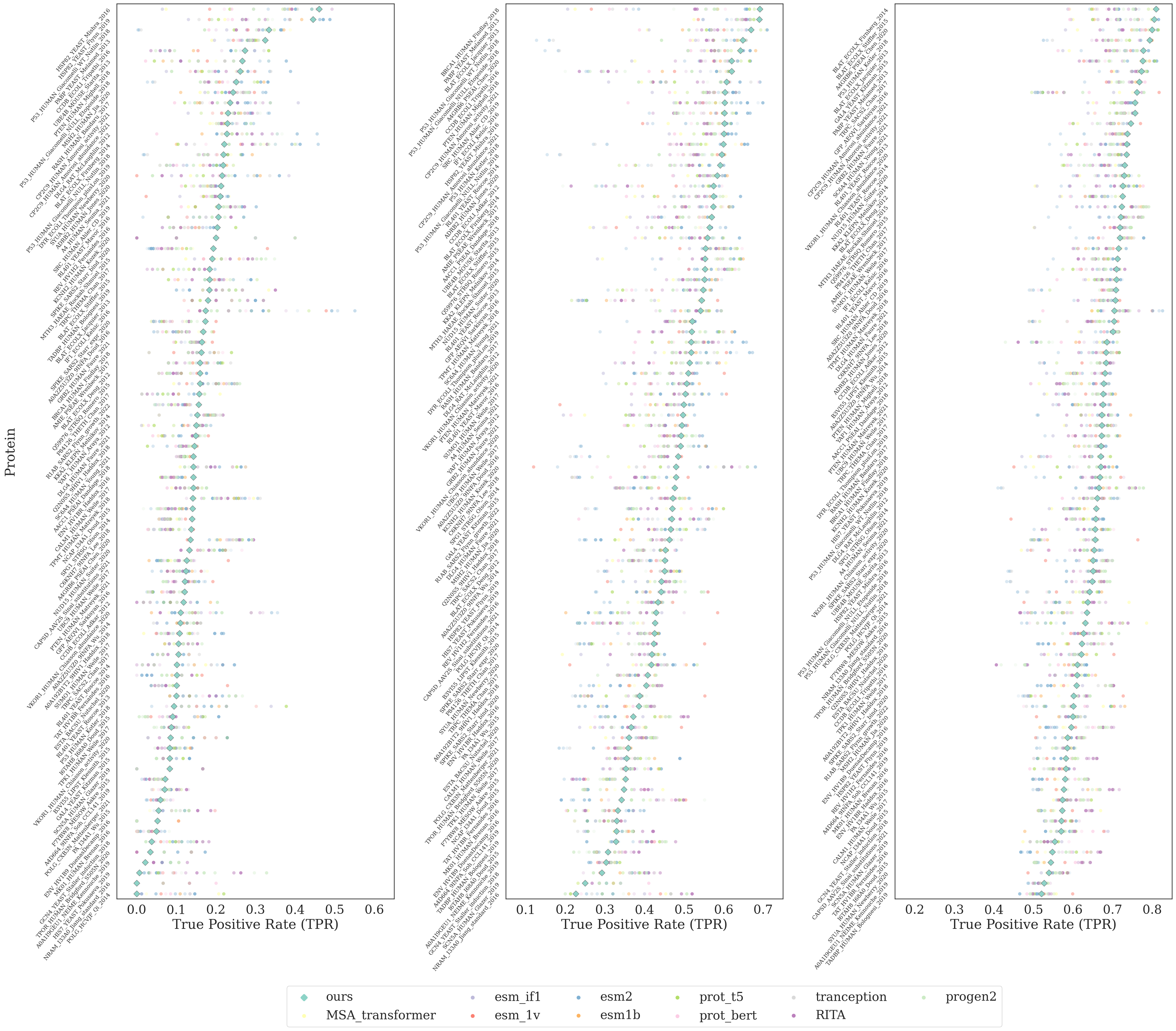}
    \caption{TPR on \textbf{ProteinGym} with top $5\%$ (left), $25\%$ (middle), and $50\%$ (right) samples be identified positive.}
    \label{fig:pg_tpr}
\end{figure}

In addition to Spearman's $\rho$ correlation, TPR scores for protein assays across all three benchmarks are presented in Figures~\ref{fig:Tm_tpr} through~\ref{fig:pg_tpr}. It should be noted that several protein assays within \textbf{DTm} and \textbf{DDG} have been subjected to a limited number of tests in the source database, \textbf{ProtThermDB}, which leads to the frequent occurrence of extreme TPR values ($0$ or $1$) at the top $5\%$ threshold. Consequently, interpretation of TPR scores is considered more meaningful at $25\%$ or $50\%$ positive rates, or at $5\%$ within the \textbf{ProteinGym} benchmark. Despite these challenges, our \plg model delivers exceptional overall performance, clearly surpassing the baseline methods on all considered test sets.

\section{Learning Curve}
% \label{sec:app:ablation}
% details, reported in tables
% We compare the learning curve for training and validation of \plg for different combinations of the number of $k$ nearest neighbors in constructing protein graphs, and the hidden dimensions. 
% In all cases, the training loss converges (we have adopted early stopping), which indicates that the learning is sufficient. The validation loss, however, varies from case to case. It can be observed that for each $k$ in $\{10,20,30\}$, there is a scenario of hidden dimension that has converges, consistent with training.

We compared the learning curve of \plg for 9 combinations of the number of $k$ nearest neighbors used in constructing protein graphs from $\{10,20,30\}$, and hidden dimension from $\{512,768,1280\}$, with respect to both training and validation. In all cases, we have used early stopping to ensure the convergence of training loss and sufficient learning. However, the validation loss varies depending on the scenario. Notably, for each value of $k$, there exists a corresponding hidden dimension scenario that converges consistently with training.

There is typically a gap between validation and training loss, indicating that there is a room of improving generalizability of the model. One possible approach to address this issue is to increase the size of both the encoder and training datasets.

\begin{figure}[t]
    \centering
    \includegraphics[width=\textwidth]{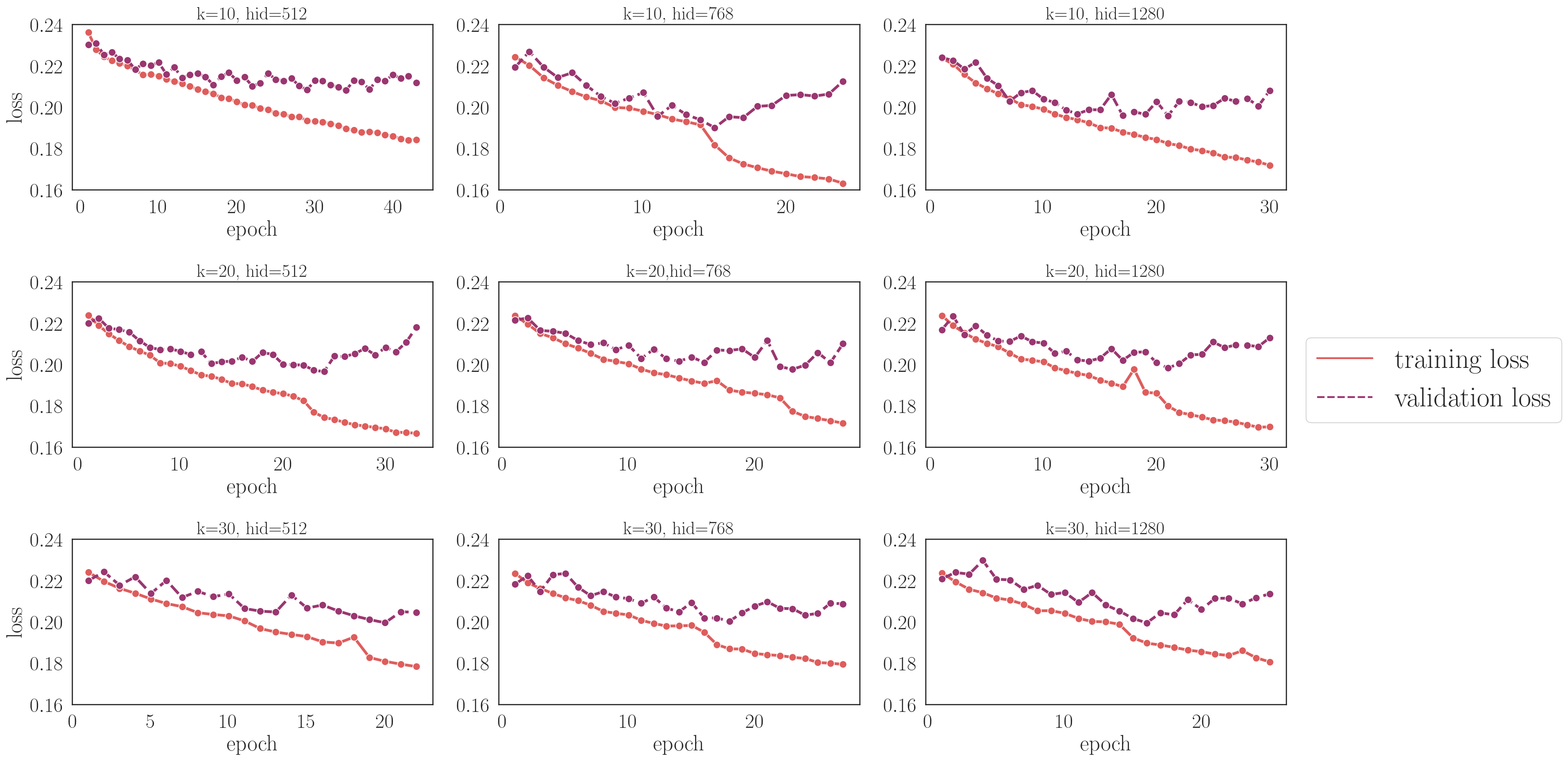}
    \caption{Training and validation losses in 9 combinations of $k=10,20,30$ nearest neighbor AAs and the hidden dimension of $512, 768, 1280$.}
    \label{fig:learnCurve}
\end{figure}

\begin{table}[!h]
\caption{Detailed Correlation Scores in Ablation Study by \textbf{ProteinGym}.}
\label{tab:ablation}
\begin{center}
\resizebox{\textwidth}{!}{
    \begin{tabular}{ccccccc}
    \toprule
    \multirow{2}{*}{Noise Encoding} & \multirow{2}{*}{Noise Rate} &
    \multirow{2}{*}{Convolution} &
    \multirow{2}{*}{ESM-2 scale (million)} &
    \multirow{2}{*}{EGNN Depth} & \multicolumn{2}{c}{\textbf{$\rho$ $\uparrow$}} \\ \cmidrule(lr){6-7}
     &  &  &  &  & Single & All \\
    \midrule
    mask & 0.15 & GCN & 650 & 6 & 0.124 & 0.210 \\
    mask & 0.15 & GAT & 650 & 6 & 0.077 & 0.177 \\
    noise & 0.4 & GCN & NA & 4 & 0.102 & 0.126 \\
    noise & 0.4 & GAT & NA & 4 & 0.070 & 0.144 \\
    noise & 0.15 & GCN & 650 & 6 & 0.103 & 0.187 \\
    noise & 0.15 & GAT & 650 & 6 & 0.101 & 0.131 \\
    noise & 0.4 & EGNN & NA & 6 & 0.311 & 0.341 \\
    mask & 0.15 & EGNN & 8 & 6 & 0.308 & 0.357 \\
    noise & 0.15 & EGNN & 8 & 6 & 0.319 & 0.366 \\
    mask & 0.15 & EGNN & 150 & 6 & 0.372 & 0.408 \\
    mask & 0.15 & EGNN & 150 & 12 & 0.370 & 0.409 \\
    noise & 0.15 & EGNN & 150 & 6 & 0.307 & 0.365 \\
    noise & 0.15 & EGNN & 150 & 12 & 0.317 & 0.364 \\
    mask & 0.15 & EGNN & 650 & 6 & 0.373 & 0.386 \\
    mask & 0.15 & EGNN & 650 & 12 & 0.403 & 0.405 \\
    mask & 0.3 & EGNN & 650 & 12 & 0.381 & 0.394 \\
    mask & 0.4 & EGNN & 650 & 12 & 0.395 & 0.403 \\
    noise & 0.15 & EGNN & 650 & 6 & 0.417 & 0.415 \\
    noise & 0.15 & EGNN & 650 & 12 & 0.342 & 0.352 \\
    noise & 0.3 & EGNN & 650 & 12 & 0.404 & 0.402 \\
    noise & 0.4 & EGNN & 650 & 12 & 0.401 & 0.401 \\
    noise & 0.25 & EGNN & 650 & 6 & 0.404 & 0.417 \\
    noise & 0.1 & EGNN & 650 & 6 & 0.409 & 0.409 \\
    noise & 0.01 & EGNN & 650 & 6 & 0.356 & 0.414 \\
    noise & 0.05 & EGNN & 650 & 6 & 0.412 & 0.356 \\
    mask & 0.15 & EGNN & 3000 & 18 & 0.057 & 0.146 \\
    noise & 0.15 & EGNN & 3000 & 18 & 0.165 & 0.227\\
    \bottomrule
    \end{tabular}
}
\end{center}
\end{table}

\section{Ablation Study}
\label{sec:app:ablation}
In Section~\ref{sec:experiment_ablation}, we conducted a comprehensive analysis of the configurations of several key components in our main algorithm. The results, which are represented by box plots in Figure~\ref{fig:ablationStudy}, amalgamate a wide array of hyper-parameter combinations. For thoroughness, we enumerate the testing scores of models associated with all these combinations in Table~\ref{tab:ablation}. Note that for one-hot encoded input node features, no \textsc{ESM-2} version was involved, we thus use \textit{NA} to indicate this scenario. For noise encoding types, we include both our denoising method (named as \textit{noise} in the table) and masked tokens (named as \textit{mask}, which refers to the last of the four noise addition strategies introduced earlier).

